# ANALYTICAL MODELS AND ESTIMATIONS FOR A MEANDER LINE TURN PROTECTING AGAINST AN ULTRAWIDE BAND PULSE

Roman S. Surovtsev [a, b], Alexander V. Nosov [a], Talgat R. Gazizov [a, b]

[a] Tomsk State University of Control Systems and Radioelectronics, Tomsk, Russian Federation,
[b] Sirius University of Science and Technology, Sochi, Russian Federation
surovtsevrs@gmail.com, alexns2094@gmail.com, talgat@tu.tusur.ru

**Abstract:** The paper presents analytical models for calculating the response to a pulse excitation in a turn of a meander line. One of the models was used to derive a condition for equalizing the voltage amplitudes for three pulses at the end of the turn which has different per-unit-length delays of even and odd modes. The obtained models and equalization condition were validated by quasistatic simulation. The results will allow for accelerated design and optimization7 of this type of structures for protection against ultrawide band pulses without costly multivariate analysis and calculation of response.

## I. INTRODUCTION

Meander lines (ML) are a common component of printed circuit boards (PCBs) in modern devices. Their traditional purpose is to delay the signals coming from different points of the PCB for synchronizing them at the receiving point. Also, MLs have found an application in receiving and transmitting antennas [1], inductors [2], multilayer capacitors [3], filtering devices [4], devices for correcting group delay time and phase correctors [5].

Most studies of MLs are aimed to minimize useful signal distortions arising from cross-couplings. Since, in its first approximation, an ML can be represented by a set of pairs of coupled lines, the numerical methods developed for coupled lines are used to analyze ML parameters. However, the use of these methods is not always advisable. For example, in a number of special cases (with negligible loss and dispersion), when the computational costs of simulation by a numerical method are high (especially in case of optimization), it is reasonable to use analytical models for quick estimates, which also provide acceptable accuracy. So, for example, the approach for analyzing crosstalk, signal propagation delay and pulse distortions in interconnects is well known [6]. Expressions in a closed form are presented for determining the transfer functions of $N$ coupled interconnects with arbitrary impedances at the ends [7]. Also noteworthy are models based on the numerical inverse Laplace transform [8, 9], and analytical models for periodic multistage structures of single and coupled lines [10]. Particularly, the analytical models of [10] have been successfully modified in [11, 12].

One of the new applications of MLs is the protection against an ultrashort pulse (UPS) by its decomposition into pulses with equal and lower amplitudes: into 2 main pulses in an air line [13] and into 3 ones in a microstrip line [14]. In the noted studies, the analysis was performed in the frequency domain based on algorithmic mathematical models. Meanwhile, the use of such models is costly, especially if it is necessary to optimize the lines in order to equalize and minimize the amplitudes of the decomposed pulses [15]. Therefore, it is important to obtain simple analytical models for quick and a priori estimates of the pulse amplitudes at the end of the ML turn and also the conditions for their equalization. The aim of this paper is to do this for an arbitrary ML turn which has different per-unit-length delays of even and odd modes.

## II. INITIAL DATA

Figure 1 shows a circuit diagram of the ML under investigation. It consists of a reference and two signal parallel conductors with the length $l$, interconnected at one end. One of the signal conductors is connected to the e.m.f. source $E$ with internal admittance $Y_0$, while another – to the load $Y_0$. Since the signal at the ML end is represented as a sequence of main pulses, first it is necessary to analytically obtain the amplitude of each of the pulses. For this, it is convenient to use analytical models for one segment of a symmetric coupled transmission line obtained on the basis of models for a single line with admittance $Y_1$ and per-unit-length delay $\tau_1$ (see Figure 2) [10].

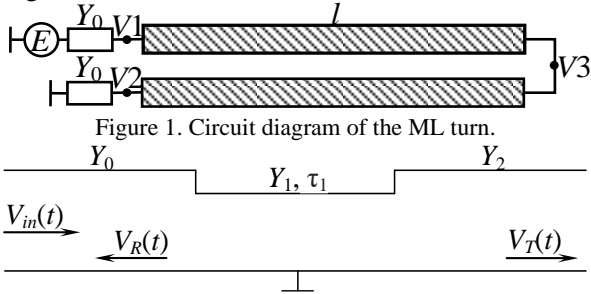

Figure 1. Circuit diagram of the ML turn.

Figure 2. Equivalent circuit of the transmission line segment with the terminations.

To make further discourse clear, we will first present these models. The response components at the far end (($V_0(t)$ considering the transmitted wave) and near end ($V_1(t)$ and $V_1(t)$ considering the reflections from the beginning and end of the line segment) of the structure are [10]:

$$V_0(t) = \frac{2Y_0}{Y_0+Y_1}\frac{2Y_1}{Y_1+Y_2}V_{in}(t-l\tau_1),$$

$$V_1'(t) = \frac{Y_0-Y_1}{Y_0+Y_1}V_{in}(t), \quad (1)$$

$$V_1''(t) = \frac{2Y_0}{Y_0+Y_1}\frac{2Y_1}{Y_0+Y_1}\frac{Y_1-Y_2}{Y_2+Y_1}V_{in}(t-2l\tau_1).$$

The following expressions, when the even or odd numbers of reflections $k_{ref}$ are set, allow us to calculate the response at the far (which consider the components that experienced an even number of reflections) or near (which consider the components that experienced an odd number of reflections) ends of the line [10]:

$$V_T(t) = V_0(t) + \sum_{k=1}^{k_{ref}/2} V_{outK}(t),$$

$$V_R(t) = V_1'(t) + V_1''(t) + \sum_{k=1}^{(k_{ref}-1)/2} V_{refK}(t) \quad (2)$$

where

$$V_{outK}(t) = \frac{2Y_0}{Y_0+Y_1}\frac{2Y_1}{Y_1+Y_2}$$
$$\times V_{in}(t-(2k+1)l\tau_1)\prod_{i=1}^{k}\frac{Y_1-Y_2}{Y_1+Y_2}\frac{Y_1-Y_0}{Y_0+Y_1},$$

$$V_{refK}(t) = \frac{2Y_0}{Y_0+Y_1}\frac{2Y_1}{Y_0+Y_1}\frac{Y_1-Y_2}{Y_1+Y_2}$$
$$\times V_{in}(t-2(k+1)l\tau_1)\prod_{i=1}^{k}\frac{Y_1-Y_2}{Y_1+Y_2}\frac{Y_1-Y_0}{Y_0+Y_1}.$$

Expressions (1)–(2) for a single line are applicable for a symmetrical (in cross-section and on loads) coupled line if we write them separately replacing index "1" in $Y_1$ and $\tau_1$ with indices "$e$" and "$o$" for the even and odd modes, respectively. Then, when the voltage at the beginning of the active conductor $V_{in}(t)$ is equal to half the exciting e.m.f., the response components for the even and odd modes can give the responses at each node of the coupled line [10]

$$V_1(t) = \frac{1}{2}\left[V_R^e(t) + V_R^o(t)\right],$$

$$V_2(t) = \frac{1}{2}\left[V_R^e(t) - V_R^o(t)\right],$$

$$V_3(t) = \frac{1}{2}\left[V_T^e(t) + V_T^o(t)\right], \quad (3)$$

$$V_4(t) = \frac{1}{2}\left[V_T^e(t) - V_T^o(t)\right],$$

where $V_1(t)$ and $V_3(t)$ are the responses at the beginning and end of the active conductor, and $V_2(t)$ and $V_4(t)$ - of the passive conductor.

### III. MODELS FOR CALCULATING THE TIME RESPONSE IN THE ML TURN

The structure in Figure 1 is a coupled line with conductors without terminations and interconnected at the far end. Then, for the far end (in Figure 2 and in (1–2)) $Y_2=\infty$ for the odd mode and $Y_2=0$ for the even mode. Therefore, the expressions (1–2) for each mode will be greatly simplified. Using (3), we obtain the final expressions for calculating the responses at nodes $V1$, $V2$, $V3$:

$$V_1(t) = \frac{V_{in}(t)}{2}\left(\frac{Y_0-Y_e}{Y_0+Y_e} + \frac{Y_0-Y_o}{Y_0+Y_o}\right) + 2Y_0\left(\frac{Y_e V_{in}(t-2l\tau_e)}{(Y_e+Y_0)^2}\right.$$
$$\left. - \frac{Y_o V_{in}(t-2l\tau_o)}{(Y_o+Y_0)^2}\right) + 2Y_e Y_0 \sum_{i=2}^{k_{ref}}(-1)^{i+1}V_{in}(t-2l\tau_e i)$$
$$\times (Y_e+Y_0)^{-(1+i)}(Y_0-Y_e)^{i-1} - 2Y_o Y_0 \quad (4)$$
$$\times \sum_{i=2}^{k_{ref}} V_{in}(t-2l\tau_o i)\times(Y_o+Y_0)^{-(1+i)}(Y_0-Y_o)^{i-1},$$

$$V_2(t) = \frac{V_{in}(t)}{2}\left(\frac{Y_0-Y_e}{Y_0+Y_e} - \frac{Y_0-Y_o}{Y_0+Y_o}\right) + 2Y_0\left(\frac{Y_e V_{in}(t-2l\tau_e)}{(Y_e+Y_0)^2}\right.$$
$$\left. + \frac{Y_o V_{in}(t-2l\tau_o)}{(Y_o+Y_0)^2}\right) + 2Y_e Y_0 \sum_{i=2}^{k_{ref}}(-1)^{i+1}V_{in}(t-2l\tau_e i) \quad (5)$$
$$\times (Y_e+Y_0)^{-(1+i)}(Y_0-Y_e)^{i-1} + 2Y_o Y_0$$
$$\times \sum_{i=2}^{k_{ref}} V_{in}(t-2l\tau_o i)(Y_o+Y_0)^{-(1+i)}(Y_0-Y_o)^{i-1},$$

$$V_3(t) = V_4(t) = \frac{2Y_0 V_{in}(t-l\tau_e)}{(Y_e+Y_0)} + 2Y_0 \quad (6)$$
$$\times \sum_{i=2}^{k_{ref}}(-1)^{i+1}V_{in}(t-l\tau_e(2i-1))(Y_e+Y_0)^{-i}(Y_0-Y_e)^{i-1}.$$

To verify the expressions (4–6), we calculated the time responses at nodes $V1$, $V2$, $V3$ to the USP excitation (Figure 3) in the TALGAT

software by a numerical method in the frequency domain [16]. The parameters of the ML cross-section, the source and load are taken from [14]. The obtained waveforms completely coincide.

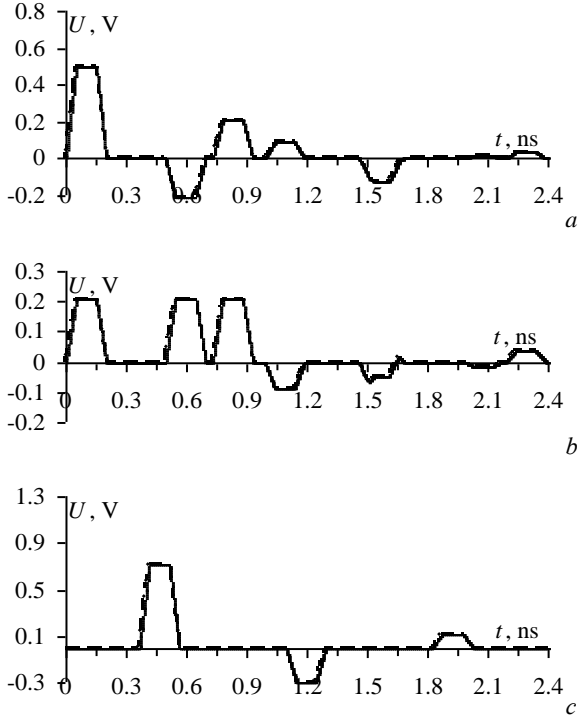

Figure 3. Voltage waveforms at nodes $V1$ ($a$), $V2$ ($b$), $V3$ ($c$) of the circuit in Figure 1$a$, calculated using expressions (4–6) (- -) and with the TALGAT software (—).

## IV. CONDITION FOR EQUALIZING THE PULSE AMPLITUDES AT THE END OF THE MEANDER LINE

From (5) it is easy to get the expressions of the normalized amplitudes of first (crosstalk – $V_c$), second (odd mode – $V_o$) and third (even mode – $V_e$) pulses

$$V_c = \frac{Y_0(Y_o - Y_e)}{(Y_o + Y_0)(Y_e + Y_0)}, \quad V_o = \frac{2Y_0 Y_o}{(Y_o + Y_0)^2}, \quad V_e = \frac{2Y_0 Y_e}{(Y_e + Y_0)^2}. \quad (7)$$

Since the amplitudes of the second and third pulses are equal, $V_e = V_o$, we obtain

$$Y_0 = \sqrt{Y_e Y_o}. \quad (8)$$

From the condition of equality of the first and second pulses amplitudes $V_c = V_o$, we obtain

$$\frac{Y_o - 3Y_e}{Y_0} + \frac{Y_e}{Y_o} = 1. \quad (9)$$

Considering (8) and substituting $k = \sqrt{(Y_o/Y_e)}$, which has the physical meaning of the coupling coefficient, we obtain the cubic equation

$$k^3 - k^2 - 3k - 1 = 0$$

having one physical root $k = \sqrt{2} + 1 \approx 2.413$ which determines the normalized amplitude at the end as $(k-1)/(k+1) = 1/(1+\sqrt{2}) \approx 0.414$.

To verify the obtained analytical models and estimations we calculated matrices and the time responses at nodes $V1$ and $V2$ to the USP excitation in the TALGAT software. The matrices of electrostatic (**C**) and electromagnetic (**L**) inductance coefficients, and characteristic impedance matrix (**Z**) are the following:

$$\mathbf{C} = \begin{bmatrix} 232.06 & -138.12 \\ -138.12 & 232.06 \end{bmatrix} \text{pF/m},$$

$$\mathbf{L} = \begin{bmatrix} 390.34 & 309.03 \\ 309.03 & 390.34 \end{bmatrix} \text{nH/m},$$

$$\mathbf{Z} = \begin{bmatrix} 50.5516 & 35.7304 \\ 35.7304 & 50.5516 \end{bmatrix} \text{Ohm}.$$

We calculated the impedances of the odd and even modes from the matrix **Z** ($Z_o$=14.8211 Ohm and $Z_e$=86.282 Ohm), which ensure the condition $k = \sqrt{Z_e/Z_o} \approx 2.413$. Figure 4 shows the waveforms at nodes $V1$ and $V2$. The first 3 pulses at the end of the turn have the same amplitudes, equal to 41.5% of the input level. Thus, the obtained models and estimates were confirmed.

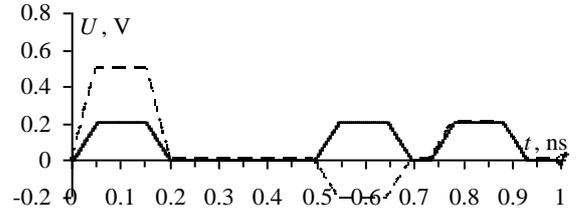

Figure 4. Voltage waveforms at nodes $V1$ (– –) and $V2$ (—) of the microstrip ML turn when $k \approx 2.413$.

## V. CONCLUSION

Analytical models for calculating the response to a time domain excitations of an arbitrary ML turn were obtained and verified. The models made it possible to obtain the condition (in particular case of matching) for the equality of pulse amplitudes after the USP decomposition at the end of the ML turn. This will further allow us to perform accelerated design and parametric optimization of such structure without costly multivariate analysis and calculation of the response. The obtained analytical models and estimations were validated by quasistatic simulation.

## ACKNOWLEDGMENT

The research was supported by the Russian Foundation for Basic Research (Project 19-37-51017).

The authors are with the Tomsk State University of Control Systems and Radioelectronics, 634050, Tomsk, Russia (e-mail: surovtsevrs@gmail.com; alexns2094@gmail.com; talgat@tu.tusur.ru). T.R. Gazizov and R.S. Surovtsev are also with Sirius University of Science and Technology, 354340, Sochi, Russia.